\documentclass[conference,a4paper]{IEEEtran}
\IEEEoverridecommandlockouts

\usepackage{hyperref}
\usepackage[cmex10]{amsmath}%American Math Society(AMS) math formatting
\usepackage{amssymb,amsfonts}%AMS extra symbols and fonts
\usepackage{dblfloatfix}%fix double column figure ordering and placement

\usepackage[ruled,vlined]{algorithm2e}
\usepackage{graphicx}
\graphicspath{{Figures/PDF/}{Figures/PNG/}}

\usepackage{booktabs}
\usepackage{siunitx}
\usepackage[numbers,compress]{natbib}
\usepackage{texnames}
\usepackage{bm,bbm}
\usepackage{orcidlink}

\begin{document}
\title{\uppercase{Forward-Scatter Bistatic RCS of a PEC Sphere: A Mie-Theory Analysis\vspace{-7pt}}}

\author{
	\IEEEauthorblockN{Khalid El-Darymli, \textit{SMIEEE} and Christoph H. Gierull, \textit{SMIEEE}}
	\IEEEauthorblockA{Department of National Defence, Defence Research and Development Canada, Ottawa, ON K1A 0Z4\\
		\{khalid.el-darymli, christoph.gierull\}@drdc-rddc.gc.ca
		\vspace{-7pt}}
}

\maketitle

\begin{abstract}Forward-scatter radar (FSR) represents an extreme bistatic configuration in which the transmitter--target--receiver geometry approaches collinearity, yielding a bistatic angle near $180^\circ$. In this regime, diffraction-dominated scattering can produce bistatic radar cross sections (RCS) that substantially exceed their monostatic counterparts, a property that underpins FS and passive bistatic radar concepts. This paper presents a rigorous electromagnetic study of the bistatic RCS of a perfectly electrically conducting (PEC) sphere using exact Lorenz--Mie theory. The analysis addresses two practical objectives: (i) quantifying the sensitivity of FS enhancement to deviations from the ideal $180^\circ$ bistatic angle as a function of electrical size, and (ii) decomposing the FS response into the canonical Lorenz--Mie polarization channels $S_1$ and $S_2$, thereby establishing a polarization-resolved benchmark. The results, however, are governed by the dimensionless size parameter $ka$ and therefore generalize directly to other frequencies and bistatic configurations. By explicitly relating full-wave simulation results to Babinet-based FS scaling and optical-theorem interpretations in the high-frequency limit, the paper provides a physically transparent benchmark for FS and passive bistatic radar analysis.
\end{abstract}


\section{Introduction}\label{sec:1}
Forward-scatter radar (FSR) is a special case of bistatic radar in which the transmitter and receiver are positioned on opposite sides of the target, such that the bistatic angle $\beta$ approaches $180^\circ$ \cite{naka2000distant, Cherniakov2007, Falconi2017}. This geometry is particularly relevant to passive coherent location (PCL) and illuminators-of-opportunity scenarios, where opportunistic transmitter--receiver placements may naturally produce near-FS conditions \cite{Gashinova2013,NATO_PassiveRadar}. A distinctive feature of FSR is that the dominant contribution to the received field arises from \emph{diffraction} by the target's geometric shadow rather than specular reflection, often yielding large effective RCS even for low-observable targets \cite{Willis2007,Glaser1985}. Recent work continues to advance passive bistatic processing under diverse illuminators and waveforms \cite{Chen2024_Sensors_FH, 9826859, Paine2024_IET_PRMP_Ch3}.

The physical interpretation of FS is closely tied to Babinet's principle and the optical theorem. In the high-frequency (optical) limit, the FS response of an opaque object can be related to its projected shadow area, and the main-lobe angular width is inversely proportional to target extent \cite{Falconi2017}. These properties are attractive for detection, but they also imply that FS enhancement is confined to a narrow angular region around $\beta = 180^\circ$, raising a practical question: \emph{how close to $180^\circ$ must the bistatic angle be to retain ``acceptable'' gain?} This question is also central to FS detection and estimation methods proposed in recent literature \cite{Falconi2017, rs16020211}.

\textbf{Contributions.} This paper provides a rigorous benchmark study using exact Lorenz--Mie theory for a PEC sphere. Specifically, we:
\begin{enumerate}
	\item compute polarization-resolved bistatic RCS via the Mie scattering amplitudes $S_1(\beta)$ and $S_2(\beta)$ and visualize the full angular dependence (Figs.~\ref{fig:beta90}--\ref{fig:avg2d});
	\item connect FS scaling to Babinet/diffraction physics using two practical FS relations (Fig.~\ref{fig:beta180}): a peak FS approximation $\sigma_{\mathrm{FS(\max)}}$ and a FS lobe-width estimate $\theta_{\mathrm{FS}}$;
	\item quantify sensitivity to deviations from $\beta=180^\circ$ and discuss the trade-off between angular misalignment and FS gain, motivated by realistic bistatic and passive radar geometries.
\end{enumerate}

\par

\par
Electromagnetic scattering behavior is commonly classified according to the dimensionless size parameter $ka = 2\pi a/\lambda$, where $a$ denotes a characteristic target dimension (the sphere radius in this paper) and $k$ is the free-space wavenumber.  This parameter delineates three canonical regimes in the literature. For $ka < 1$, the scattering lies in the \emph{Rayleigh regime}, where the target is electrically small and the response is quasi-static and smoothly varying. For $1 \lesssim ka \lesssim 10$, the scattering enters the \emph{Mie (resonance) regime}, characterized by strong full-wave effects, oscillatory behavior, and sensitivity to geometry, polarization, and bistatic angle. For $ka > 10$, the scattering transitions into the \emph{optical (geometric-optics) regime}, where diffraction, shadowing, and specular mechanisms dominate and FS behavior is well described by Babinet’s principle and the optical theorem. In this paper, the boundaries $ka = 1$ and $ka = 10$ are adopted as practical demarcations between these regimes, consistent with standard electromagnetic scattering literature, and are explicitly highlighted in Figs.~\ref{fig:beta90}--\ref{fig:beta180} to guide interpretation of the bistatic RCS behavior.

The remainder of this paper is organized as follows. 
Section~\ref{sec:2} defines the bistatic geometry, polarization bases, and notation used throughout the paper. 
Section~\ref{sec:3} summarizes the Lorenz--Mie formulation for a PEC sphere and introduces the bistatic RCS definitions for the $S_1$ and $S_2$ polarization channels. 
Section~\ref{sec:4} relates FS behavior to Babinet-diffraction scaling and introduces two practical FS relations used to interpret the numerical results. 
Section~\ref{sec:5} outlines the numerical methodology used to compute polarization-resolved and averaged bistatic RCS. 
Section~\ref{sec:6} presents the simulation results and analyzes the sensitivity of FS enhancement to deviations from $\beta = 180^\circ$, including polarization effects and near-forward angular tolerance. 
Section~\ref{sec:7} discusses the implications of the results for practical bistatic and passive radar geometries, including an operational-scale interpretation based on Satellite Digital Audio Radio Service (SDARS) emitters that DRDC uses with passive radar, also in FS configurations, exemplified by the SXM-8 (B) system. Finally, Section~\ref{sec:8} concludes the paper.

\section{Geometry and Symbols}\label{sec:2}
\subsection{Bistatic configuration and bistatic angle}
Fig.~\ref{fig:geom} defines the bistatic scattering geometry for a PEC sphere of radius $a$ centered at the origin. A monochromatic plane wave with wavelength $\lambda$ and wavenumber $k=2\pi/\lambda$ illuminates the sphere along the incident wave vector
\begin{equation}
	\mathbf{k}_i = k\,\hat{\mathbf{z}}.
\end{equation}
The scattered field is observed in a direction defined by the scattered wave vector $\mathbf{k}_s$. In Fig.~\ref{fig:geom}, two representative scattered directions are shown: a near-FS direction $\mathbf{k}_{s_1}$ and a general bistatic direction $\mathbf{k}_{s_2}$. The \emph{bistatic angle} $\beta$ is the angle between incident and scattered propagation directions:
\begin{equation}
	\beta \triangleq \cos^{-1}\!\left(\hat{\mathbf{k}}_i \cdot \hat{\mathbf{k}}_s\right),\qquad 0^\circ \le \beta \le 180^\circ.
	\label{eq:beta_def}
\end{equation}
Accordingly, $\beta=0^\circ$ corresponds to backscatter (monostatic direction), $\beta=90^\circ$ to side scatter, and $\beta=180^\circ$ to ideal FS.
\subsection{Scattering plane and polarization bases}
\begin{figure}[t]
	\centering
	\includegraphics[width=170pt]{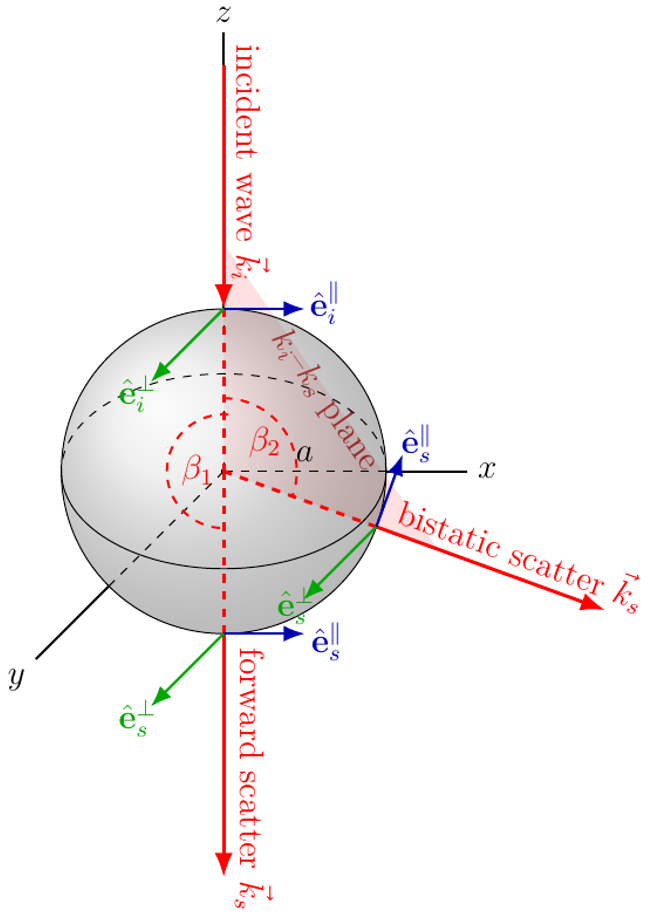}\vspace{-7pt}
	\caption{Bistatic scattering geometry and symbols for a PEC sphere of radius $a$. The incident plane wave propagates along $\mathbf{k}_i$ and is scattered along $\mathbf{k}_s$; the bistatic angle $\beta$ is defined by Eq.~\eqref{eq:beta_def}. The scattering plane is spanned by $\mathbf{k}_i$ and $\mathbf{k}_s$. Polarization bases are defined as $\parallel$ (in-plane) and $\perp$ (out-of-plane): $\hat{\mathbf{e}}_{i}^{\parallel},\hat{\mathbf{e}}_{i}^{\perp}$ for incidence and $\hat{\mathbf{e}}_{s}^{\parallel},\hat{\mathbf{e}}_{s}^{\perp}$ for scattering.}\vspace{-10pt}
	\label{fig:geom}
\end{figure}

The \emph{scattering plane} (shaded in Fig.~\ref{fig:geom}) is spanned by $\mathbf{k}_i$ and $\mathbf{k}_s$. Polarization is decomposed into unit vectors \emph{parallel} ($\parallel$) and \emph{perpendicular} ($\perp$) to this plane. The incident polarization basis is $\{\hat{\mathbf{e}}_{i}^{\parallel},\hat{\mathbf{e}}_{i}^{\perp}\}$ and the scattered basis is $\{\hat{\mathbf{e}}_{s}^{\parallel},\hat{\mathbf{e}}_{s}^{\perp}\}$, as labeled in Fig.~\ref{fig:geom}. For a sphere, azimuthal symmetry implies that cross-polarization is absent; the two canonical polarization channels remain decoupled and are fully characterized by the Mie scattering amplitudes $S_1(\beta)$ and $S_2(\beta)$ \cite{Bohren1983}.

\section{Lorenz--Mie Formulation for a PEC Sphere}\label{sec:3}
\subsection{Scattering amplitudes and bistatic RCS}
In the far field, the scattered electric field of a sphere can be expressed in terms of two scalar scattering amplitudes, $S_1(\beta)$ and $S_2(\beta)$ \cite{Bohren1983,Matzler2002, Brock2016}. Using the standard polarization-resolved bistatic RCS formulation,
\begin{equation}
	\sigma_{\perp}(\beta)=\frac{4\pi}{k^2}\left|S_1(\beta)\right|^2,\qquad
	\sigma_{\parallel}(\beta)=\frac{4\pi}{k^2}\left|S_2(\beta)\right|^2.
	\label{eq:rcs_S1S2}
\end{equation}
An unpolarized (or polarization-averaged) bistatic RCS is formed as
\begin{equation}
	\sigma_{\mathrm{avg}}(\beta)=\tfrac{1}{2}\left(\sigma_{\perp}(\beta)+\sigma_{\parallel}(\beta)\right).
	\label{eq:rcs_avg}
\end{equation}

\subsection{PEC sphere coefficients}
For a sphere, $S_1$ and $S_2$ are given by Mie series involving spherical Bessel and Hankel functions and the Mie coefficients $a_n$ and $b_n$ \cite{Bohren1983}. For a PEC sphere, these coefficients simplify (compared to penetrable media), and the series truncation order can be chosen based on the size parameter $x=ka$ (e.g., $n_{\max}\approx x+4x^{1/3}+2$ is a common guideline) \cite{Bohren1983,Matzler2002}. The numerical implementation follows these standard practices, enabling high-dynamic-range evaluation near FS.
\section{FS Scaling and Babinet Principle}\label{sec:4}
FS is diffraction-dominated and can be interpreted using Babinet's principle together with the optical theorem \cite{Falconi2017,Gashinova2013}. In the optical limit for an opaque object, the peak FS RCS is commonly approximated by the relation overlaid in Fig.~\ref{fig:beta180}:
\begin{equation}
	\sigma_{\mathrm{FS(\max)}} \approx 4\pi\left(\frac{A}{\lambda}\right)^2,
	\label{eq:sigma_fsmax_fig4}
\end{equation}
where $A$ denotes the projected geometric shadow area of the target normal to the incident propagation direction (i.e., the area of the opaque silhouette seen by the incident wave). For a sphere, this projected area reduces to the area under the circle, given by $A = \pi a^2$. The same FS model used in Fig.~\ref{fig:beta180} also expresses the \emph{angular extent} of the FS lobe by the second overlaid equation:
\begin{equation}
	\theta_{\mathrm{FS}} \approx \pi\sqrt{\frac{A}{\sigma_{\mathrm{FS(\max)}}}},
	\label{eq:theta_fs_fig4}
\end{equation}
which makes explicit the trade-off between peak enhancement and angular tolerance: increasing projected area $A$ increases $\sigma_{\mathrm{FS(\max)}}$ while narrowing $\theta_{\mathrm{FS}}$. Substituting Eq.~\eqref{eq:sigma_fsmax_fig4} into Eq.~\eqref{eq:theta_fs_fig4} yields the expected inverse scaling with electrical size,
\begin{equation}
	\theta_{\mathrm{FS}} \propto \frac{\lambda}{\sqrt{A}},
\end{equation}
and therefore, for a sphere ($A=\pi a^2$), $\theta_{\mathrm{FS}}\propto \lambda/a$. This connects the Fig.~\ref{fig:beta180} annotations directly to the practical question addressed in this paper: how much deviation from $\beta=180^\circ$ is permissible while maintaining useful FS gain.
\section{Simulation Methodology}\label{sec:5}
Bistatic RCS is computed using Lorenz--Mie theory for a PEC sphere by evaluating
Eq. \eqref{eq:rcs_S1S2} over bistatic angles and, where applicable, over a two-dimensional
angular grid to generate polarization-resolved and averaged bistatic RCS surfaces
(Figs.~\ref{fig:S1}--\ref{fig:avg2d}). The numerical workflow, including series truncation,
polarization handling, and angular sampling, is summarized in Algorithm~\ref{algo:mie_rcs}.

{\footnotesize
	\begin{algorithm}[t]
		\DontPrintSemicolon
		\SetKwInput{KwIn}{Inputs}
		\SetKwInput{KwOut}{Outputs}
		\caption{Bistatic RCS evaluation for a PEC sphere using Lorenz--Mie theory.}
		\label{algo:mie_rcs}
		\KwIn{Frequency $f$ (or wavelength $\lambda$), sphere radius $a$, bistatic angle grid $\{\beta\}$ (or 2-D angular grid), polarization channels $\{S_1,S_2\}$.}
		\KwOut{Polarization-resolved bistatic RCS $\sigma_{\perp}(\beta)$, $\sigma_{\parallel}(\beta)$, and average $\sigma_{\mathrm{avg}}(\beta)$.}
		Compute $k=2\pi/\lambda$ and size parameter $x=ka$\;
		Choose series truncation order $n_{\max}$ (e.g., based on $x$) and compute PEC Mie coefficients $\{a_n,b_n\}_{n=1}^{n_{\max}}$\;
		\ForEach{bistatic angle $\beta$ (or grid point)}{
			Evaluate Mie scattering amplitudes $S_1(\beta)$ and $S_2(\beta)$ from the truncated series\;
			Compute $\sigma_{\perp}(\beta)$ and $\sigma_{\parallel}(\beta)$ using Eq.~\eqref{eq:rcs_S1S2}\;
			Compute $\sigma_{\mathrm{avg}}(\beta)$ using Eq.~\eqref{eq:rcs_avg}\;
		}
		Generate (i) angular cuts at $\beta\in\{0^\circ,90^\circ,180^\circ\}$ (Figs.~\ref{fig:beta90}--\ref{fig:beta180}) and (ii) 3-D RCS surfaces for $S_1$, $S_2$, and $\sigma_{\mathrm{avg}}$ (Figs.~\ref{fig:S1}--\ref{fig:avg})\;
		Compare FS behavior to Eqs.~\eqref{eq:sigma_fsmax_fig4}--\eqref{eq:theta_fs_fig4}\;
	\end{algorithm}
}
\section{Results}\label{sec:6}\vspace{-2pt}
\subsection{Bistatic angle dependence}
\begin{figure}[t]
	\centering \vspace{-6pt}
	\includegraphics[width=\linewidth]{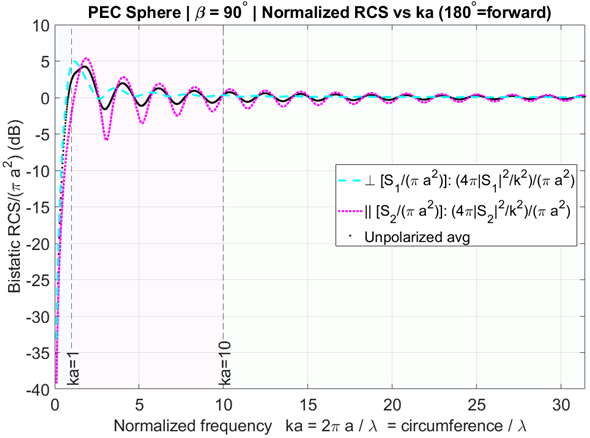}
	\caption{Bistatic RCS of a PEC sphere at $\beta=90^\circ$ (side scatter).}\vspace{-10pt}
	\label{fig:beta90}
\end{figure}

\begin{figure}[t]
	\centering
	\includegraphics[width=\linewidth]{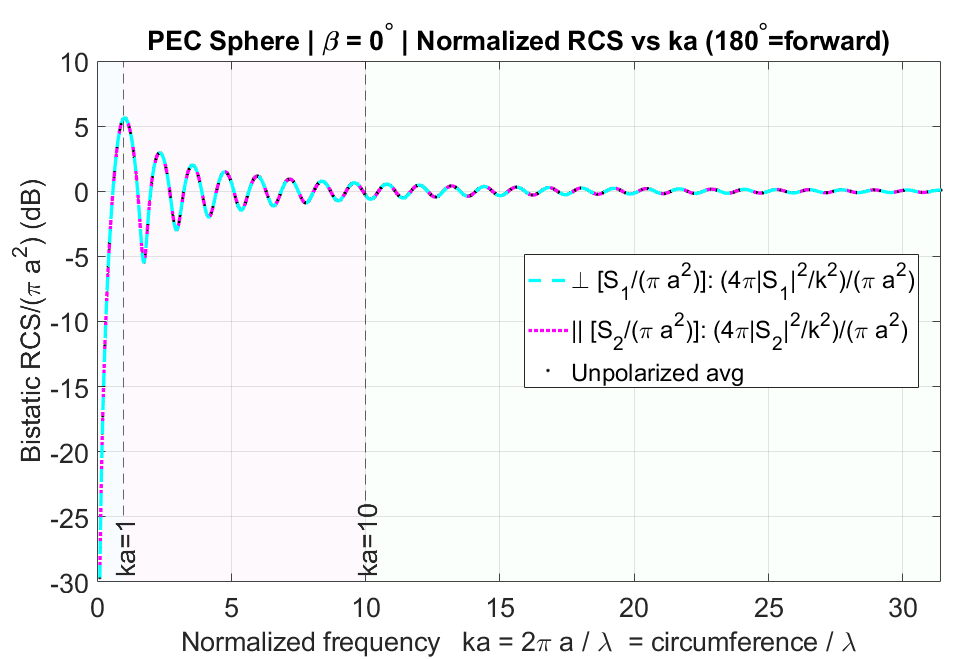}
	\caption{Bistatic RCS of a PEC sphere at $\beta=0^\circ$ (backscatter).}
	\label{fig:beta0}
\end{figure}

\begin{figure}[t]
	\centering
	\includegraphics[width=\linewidth]{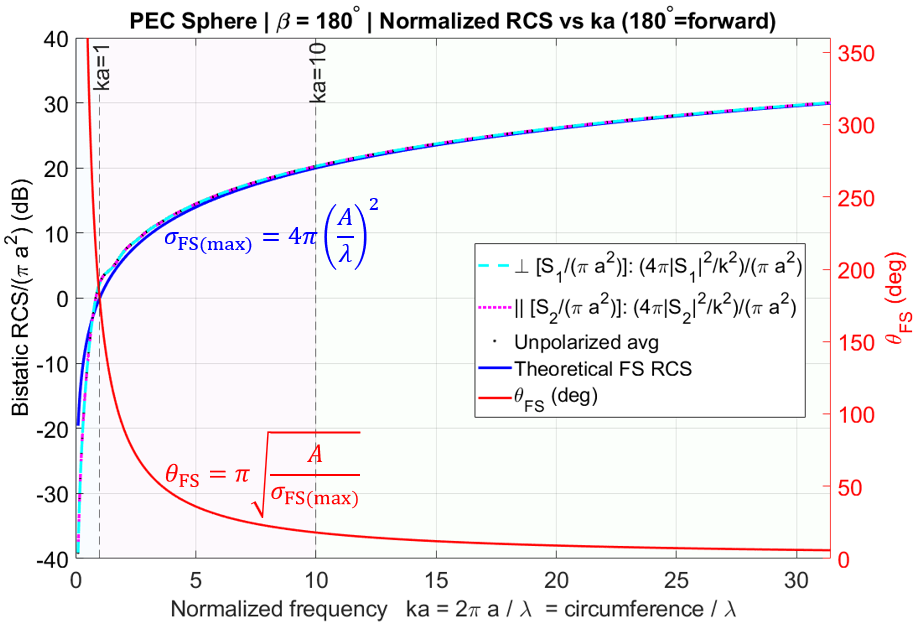}
	\caption{FS bistatic RCS at $\beta=180^\circ$; the narrow lobe motivates tolerance analysis via Eq.~\eqref{eq:theta_fs_fig4}.}
	\label{fig:beta180}
\vspace{-10pt}
\end{figure}

\vspace{-2pt}
Figs.~\ref{fig:beta90}--\ref{fig:beta180} show representative bistatic RCS cuts for $\beta=90^\circ$ (side scatter), $\beta=0^\circ$ (backscatter), and $\beta=180^\circ$ (FS). The FS case exhibits the dominant narrow lobe expected from diffraction theory and is consistent with the scaling trends predicted by Eqs.~\eqref{eq:sigma_fsmax_fig4}--\eqref{eq:theta_fs_fig4}.
\subsection{FS angular tolerance (near $\beta=180^\circ$)}\vspace{-2pt}
In practice, the bistatic angle may deviate from $180^\circ$ due to platform geometry, baseline constraints, or transmitter/receiver location uncertainty in passive scenarios. The simulations demonstrate that FS gain remains substantial within a narrow angular neighborhood of $\beta=180^\circ$, with the acceptable offset governed by the target's electrical size (via $\lambda/D$) and the corresponding lobe width Eq.~\eqref{eq:theta_fs_fig4}. In particular, offsets on the order of a few degrees (e.g., $\pm5^\circ$) can still yield strong enhancement, while larger offsets (e.g., $\pm10^\circ$) incur increasingly significant degradation as the observation direction moves off the main diffractive lobe. These trends provide actionable guidance for designing and analyzing FSR/PCL geometries.\vspace{-3pt}
\subsection{Polarization decomposition: $S_1$, $S_2$, and averaged FS response}
\begin{figure}[t]
	\centering
	\includegraphics[width=\linewidth]{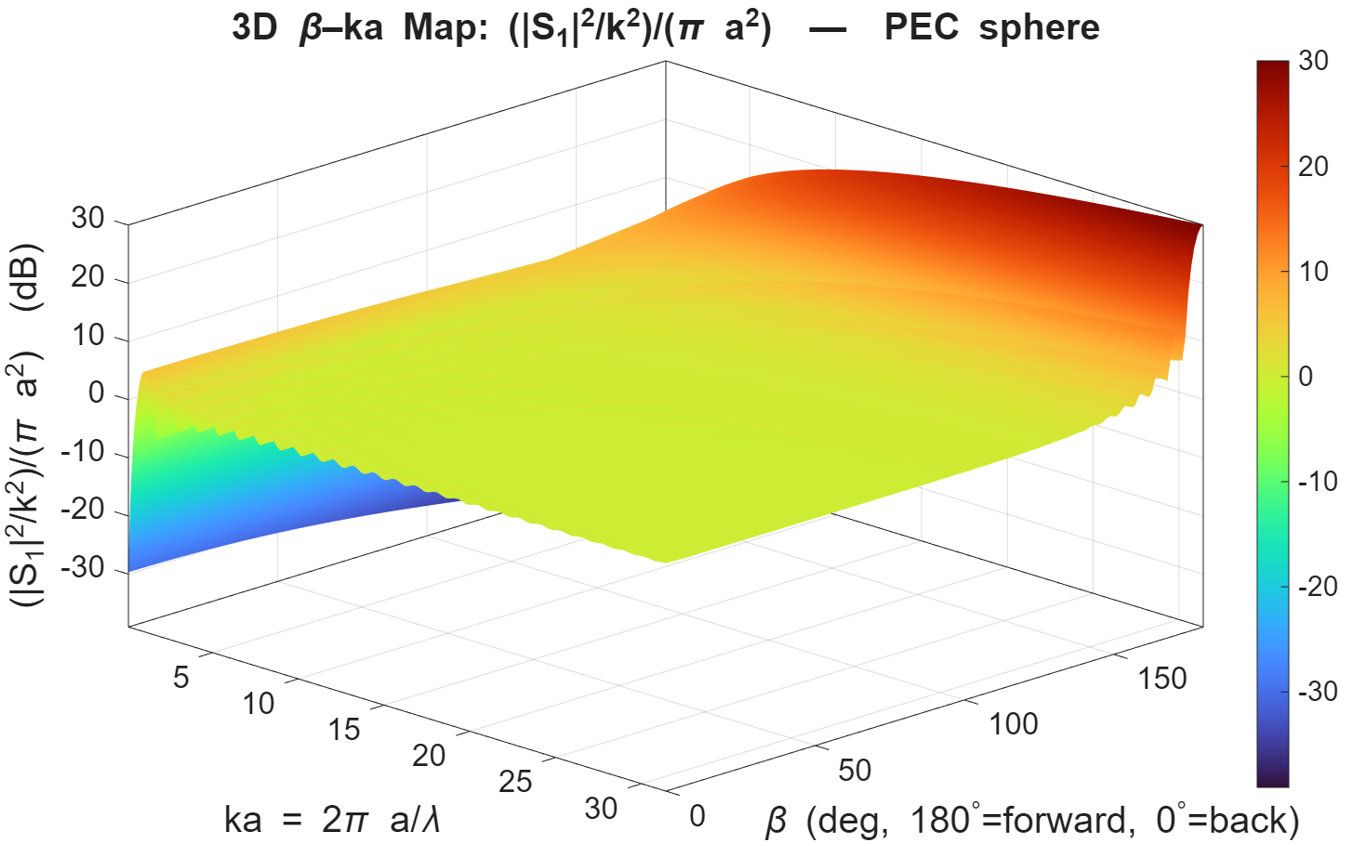}
	\caption{3-D bistatic RCS for the $S_1$ channel (perpendicular), i.e., $\sigma_{\perp}(\beta)$ in Eq.~\eqref{eq:rcs_S1S2}.}
	\label{fig:S1}
\end{figure}

\begin{figure}[t]
	\centering
	\includegraphics[width=\linewidth]{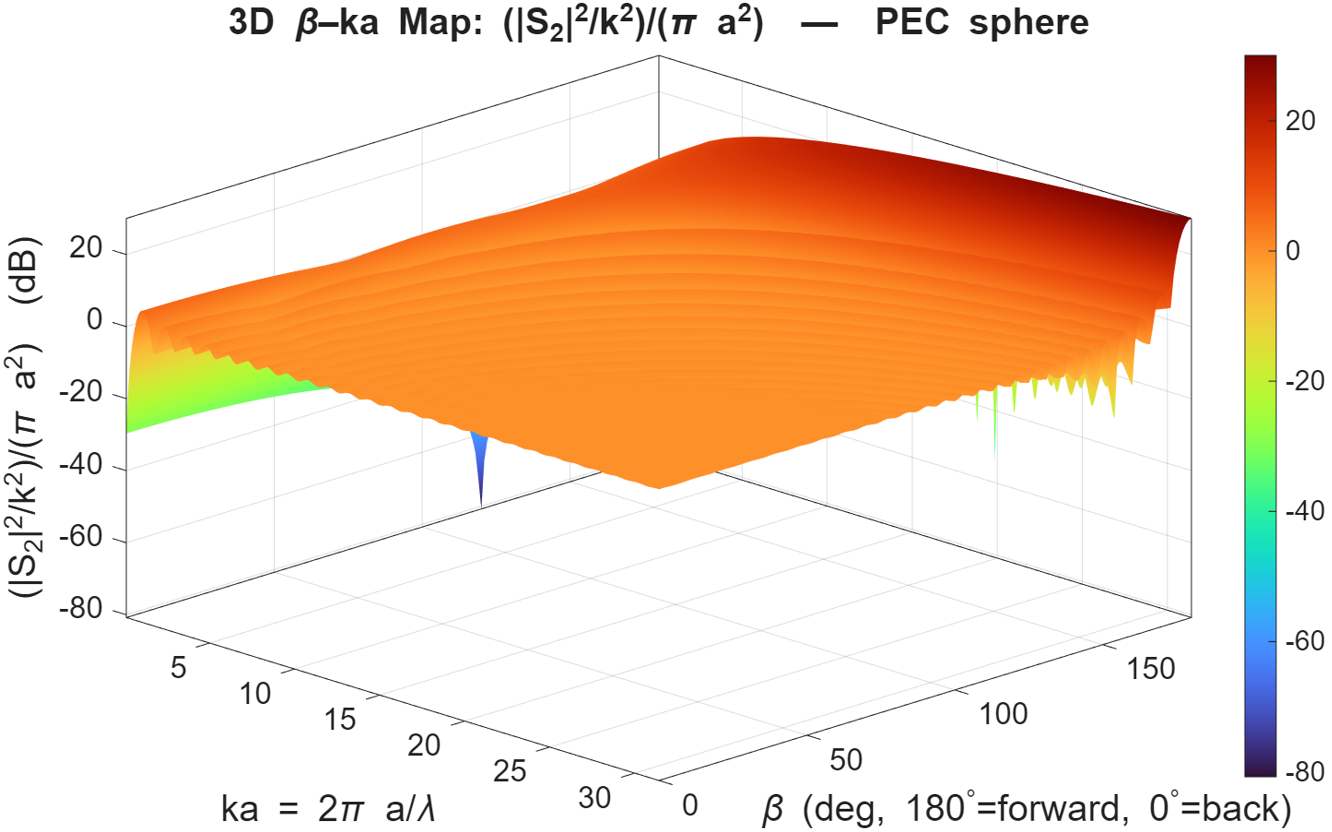}
	\caption{3-D bistatic RCS for the $S_2$ channel (parallel), i.e., $\sigma_{\parallel}(\beta)$ in Eq.~\eqref{eq:rcs_S1S2}.}
	\label{fig:S2}
	\vspace{-10pt}
\end{figure}

\begin{figure}[t]
	\centering
	\includegraphics[width=\linewidth]{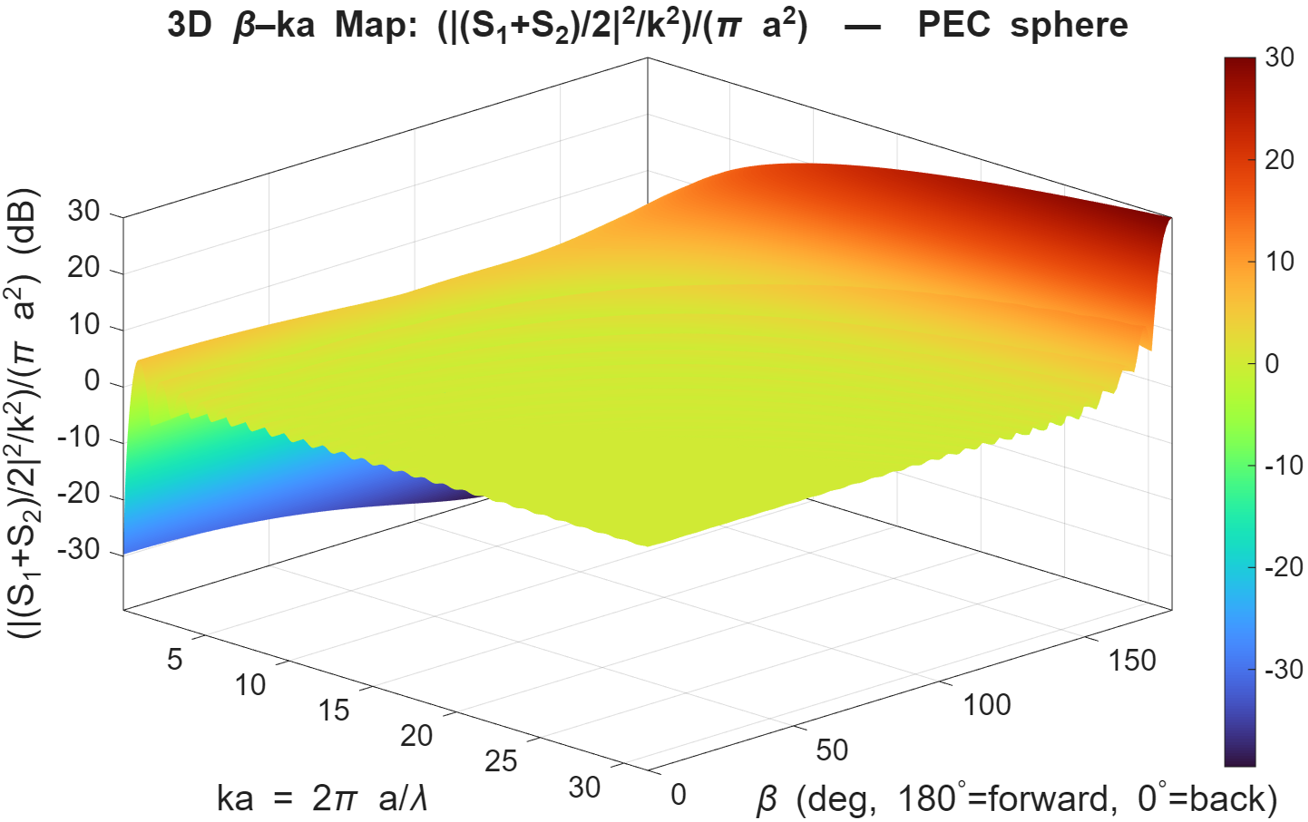}
	\caption{Polarization-averaged bistatic RCS $\sigma_{\mathrm{avg}}(\beta)$ in Eq.~\eqref{eq:rcs_avg}.}
	\label{fig:avg}
\end{figure}

\vspace{-3pt}
Figs.~\ref{fig:S1} and \ref{fig:S2} show full 3-D bistatic RCS surfaces for the two canonical polarization channels (perpendicular and parallel). For a sphere, the polarization dependence is generally modest due to symmetry, but measurable differences appear in lobe structure and sidelobe distribution. Fig.~\ref{fig:avg} shows the polarization-averaged response Eq.~\eqref{eq:rcs_avg}, which is often the appropriate metric when polarization is unknown, mixed, or when receive processing averages over polarizations.
\begin{figure}[t]
	\centering
	\includegraphics[width=0.93\linewidth]{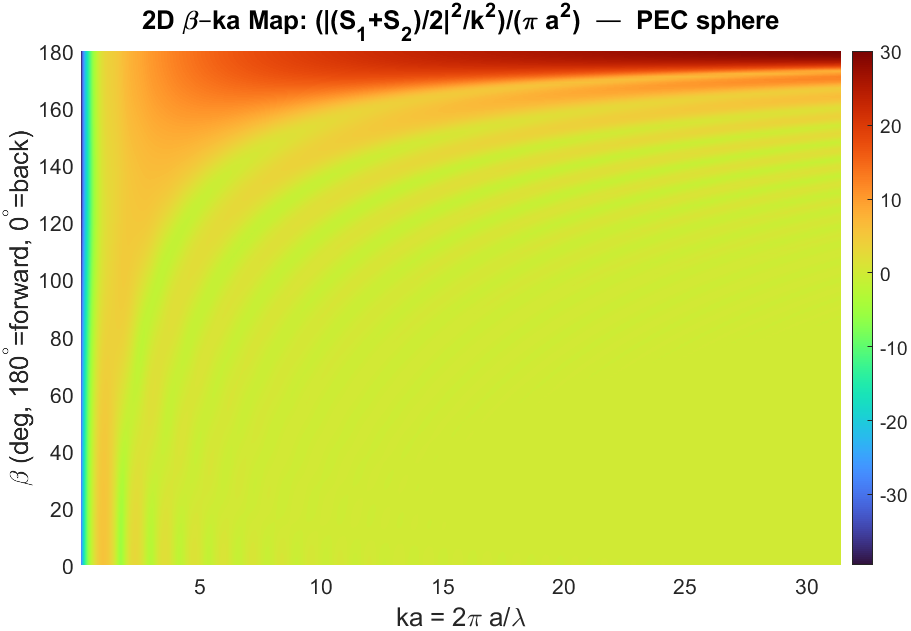}
	\caption{2-D map (from Fig.~\ref{fig:avg}) highlighting near-forward tolerance versus $ka$.}
	\label{fig:avg2d}
	\vspace{-10pt}
\end{figure}

\par
Fig.~\ref{fig:avg} summarizes the global, polarization-averaged angular structure, while Fig.~\ref{fig:avg2d} provides a compact view of the near-forward neighborhood ($\beta\approx180^\circ$) used to read off practical tolerances as a function of $ka$.
\section{Discussion}\label{sec:7}
To preserve generality, the results in Section~\ref{sec:6} are presented primarily in terms of the dimensionless size parameter $ka$ and the bistatic angle $\beta$. We now provide a concrete operational-scale interpretation by mapping the FS tolerance to an illustrative illumination frequency.

\textit{Operational scale (SXM-8 B example):} Using $f_c=2344.045~\mathrm{MHz}$ ($\lambda\approx0.128~\mathrm{m}$) as an illustrative illuminator-of-opportunity frequency \cite{igarss26_ur}, the sphere forward-lobe width can be approximated by $\theta_{\mathrm{FS}}\approx \lambda/(2a)\approx \pi/(ka)$ [rad] $\approx 180/(ka)$ [deg]. Thus, representative widths are, for example, $\theta_{\mathrm{FS}}\approx18^\circ$ at $ka=10$, $9^\circ$ at $ka=20$, $6^\circ$ at $ka=30$, and $3.6^\circ$ at $ka=50$, providing a concrete ``how-far-off-$180^\circ$'' interpretation for the near-forward ridge in Figs.~\ref{fig:beta180}, \ref{fig:avg}, and~\ref{fig:avg2d}. This frequency is used only to set physical scale; trends generalize via $ka$.
The PEC sphere provides a stringent benchmark for FS bistatic RCS analysis because it admits an exact full-wave solution. The results confirm that the \emph{maximum} FS gain occurs at $\beta=180^\circ$, consistent with diffraction-based interpretations (Babinet/optical theorem). However, because the FS lobe is narrow, the key operational question is not whether $\beta$ must be \emph{exactly} $180^\circ$, but rather whether the geometry places the receiver within the FS main lobe. The governing parameter is $\lambda/D$, which sets the lobe width according to Eq.~\eqref{eq:theta_fs_fig4}: larger spheres (or higher frequency) yield narrower lobes and therefore stricter angular tolerances, while smaller electrical sizes broaden the tolerance but reduce peak gain. Polarization differences for a sphere are secondary; nevertheless, explicitly reporting both $S_1$ and $S_2$ channels provides a clean baseline for extending the methodology to non-spherical or anisotropic targets, where polarization effects can be significant \cite{1143298,8518989,rs14030520,1268313, Larsson_Gustafsson_2021, filippini2017experimental, rodriguez2023first, 6985926, jin2024remote}.

\section{Conclusion}\label{sec:8}

We presented an exact Lorenz--Mie analysis of the bistatic RCS of a PEC sphere with emphasis on the FS regime. The enhancement is maximized at $\beta=180^\circ$ and remains useful within an angular tolerance governed primarily by $\lambda/D$, providing a rigorous reference for FS and passive bistatic radar analysis and motivating extensions to non-spherical targets.
\small
\bibliographystyle{IEEEtran}
\bibliography{references_igarss2026_updated}
\end{document}